\documentstyle[prd,aps,preprint,tighten,epsfig]{revtex}

\begin{document}
\draft

\title{Neutrino Mixing and Leptogenesis in Type-II Seesaw
Scenarios with Left-Right Symmetry}
\author{{\bf Wei Chao} \thanks{E-mail: chaowei@mail.ihep.ac.cn}, ~
{\bf Shu Luo} \thanks{E-mail: luoshu@mail.ihep.ac.cn}, ~ {\bf
Zhi-zhong Xing} \thanks{E-mail: xingzz@mail.ihep.ac.cn} \\
{\sl Institute of High Energy Physics, Chinese Academy of
Sciences, Beijing 100049, China}} \maketitle
%\date{\today}

\begin{abstract}
We propose two Type-II seesaw scenarios for the neutrino mass
matrix in the left-right symmetric model, in which the Higgs
triplet Yukawa coupling matrix takes the appealing Friedberg-Lee
texture. We show that the nearly tri-bimaximal neutrino mixing
pattern, which is especially favored by current neutrino
oscillation data, can be obtained from both scenarios. We also
show that the cosmological baryon number asymmetry can naturally
be interpreted in these two scenarios via the flavor-independent
leptogenesis mechanism.
\end{abstract}


\section{Introduction}

Current solar \cite{SNO}, atmospheric \cite{SK}, reactor \cite{KM}
and accelerator \cite{K2K} neutrino oscillation experiments have
provided us with very convincing evidence that neutrinos have
non-vanishing rest masses and their mixing involves two large
angles ($\theta^{}_{12} \sim 34^\circ$ and $\theta^{}_{23}\sim
45^\circ$) and one small angle ($\theta^{}_{13} < 10^\circ$).
These important results indicate that the standard electroweak
model, in which the gauge group is $SU(2)^{}_L \times U(1)^{}_Y$
and three neutrinos are massless Weyl particles, is actually
incomplete. There are many possibilities of extending the standard
model to accommodate massive neutrinos and to resolve or soften
other potential problems of the model itself \cite{PDG06}. One of
them, motivated by the conjecture that parity is a perfect
symmetry at high-energy scales and is spontaneously broken at
low-energy scales, is the left-right symmetric model \cite{LR}.
The elegance of this model and its testability at the LHC and ILC
experiments have recently been iterated \cite{Lee}.

The left-right symmetric model is based on the gauge group
$SU(2)^{}_L \times SU(2)^{}_R \times U(1)^{}_{B-L}$, and thus it
naturally contains both left-handed and right-handed neutrinos
together with one Higgs bi-doublet $\Phi$ and two Higgs triplets
$\Delta^{}_{L,R}$ \cite{LR}. In addition to the left-right gauge
symmetry, a discrete left-right symmetry may be introduced into
the model by requiring its invariance under
$l_L^{}\leftrightarrow (l_R^{})^c$, $q_L^{}\leftrightarrow
(q_R^{})^c$, $\Delta^{}_L \leftrightarrow \Delta^{}_R$ and $\Phi
\leftrightarrow \Phi^T$ \cite{discreteLR}. The gauge-invariant
Yukawa interactions between the fermion and Higgs sectors can be
written as
\begin{eqnarray}
- {\cal L}_Y^{} & = & \overline{q_L^{}} \Phi Y_q^{}q_R^{} +
 \overline{q_L^{}} \tilde{\Phi}\tilde{Y}_q^{} q_R^{} +
\overline{l_L^{}} \Phi  Y_l^{}l_R^{} + \overline{l_L^{}}
\tilde{\Phi} \tilde{Y}_l^{} l_R^{} \nonumber \\
&  & + ~ \frac{1}{2} \left [\overline {l_L^{}} i\tau_2^{}
\Delta_L^{}{\cal F} (l_L^{})^{c} + \overline{(l_R^{})^c}
i\tau_2^{} \Delta_R^{}{\cal F} l_R^{} \right ] + {\rm h.c.} \; ,
%----------------(1)
\end{eqnarray}
where $\tilde{\Phi} \equiv \tau_2^{}\Phi^*\tau_2^{}$ and
$(l^{}_{L,R})^c \equiv C \overline{l^{}_{L,R}}^T$ with $C$ being
the charge-conjugation matrix. As a consequence of the discrete
left-right symmetry, $Y^{}_{l,q}$ and $\tilde{Y}_{l,q}$ are both
symmetric matrices. Note that $SU(2)^{}_L \times SU(2)^{}_R \times
U(1)^{}_{B-L}$ is spontaneously broken into the standard-model
gauge group $SU(2)^{}_L \times U(1)^{}_Y$ via the vacuum
expectation value (vev) of $\Delta^{}_R$, and then the spontaneous
electroweak symmetry breaking is accomplished through the vev of
$\Phi$. Given the vevs of $\Delta^{}_L$, $\Delta^{}_R$ and $\Phi$
\cite{LR},
\begin{eqnarray}
\langle \Delta_L^{} \rangle = \left( \matrix{0 & 0 \cr v_L^{} & 0
\cr} \right) \; , \hspace{0.5cm} \langle \Delta_R^{} \rangle =
\left( \matrix{0 & 0 \cr v_R^{} & 0 \cr} \right) \; ,
\hspace{0.5cm} \langle \Phi \rangle = \left( \matrix{\kappa & 0
\cr 0 & \kappa' e^{i\alpha} \cr} \right) \; ,
%----------------(2)
\end{eqnarray}
the up-type quark, down-type quark, charged lepton and Dirac
neutrino mass matrices are
\begin{eqnarray}
M_u^{} & = & \kappa Y_q^{} + \kappa'e^{-i\alpha} \tilde{Y}_q^{} \; , \nonumber\\
M_d^{} & = & \kappa \tilde{Y}_q^{} + \kappa'e^{+i\alpha} Y_q^{} \; , \nonumber\\
M_e^{} & = & \kappa \tilde{Y}_l^{} + \kappa'e^{+i\alpha} Y_l^{} \; , \nonumber\\
M_D^{} & = & \kappa Y_l^{} + \kappa'e^{-i\alpha} \tilde{Y}_l^{} \;
.
%----------------(3)
\end{eqnarray}
Meanwhile, the left- and right-handed Majorana neutrino mass
matrices can be obtained from the corresponding mass terms in Eq.
(1) once the Higgs triplets $\Delta^{}_L$ and $\Delta^{}_R$
acquire their vevs: $M^{}_L = v^{}_L {\cal F}$ and $M^{}_R =
v^{}_R {\cal F}$. Integrating out the heavy particles (i.e., the
right-handed Majorana neutrinos and the Higgs triplets), one
obtains the effective mass matrix for three light (left-handed)
Majorana neutrinos via the Type-II seesaw relation \cite{origin}:
\begin{eqnarray}
M_\nu^{} \; \simeq \; M_L^{} - M_D^{} M_R^{-1} M_D^T \; = \;
v_L^{} {\cal F} - {1\over v_R^{}} M_D {\cal F}^{-1} M_D^T \; .
%----------------(4)
\end{eqnarray}
The phenomenon of lepton flavor mixing, which has clearly shown up
in both solar and atmospheric neutrino oscillations, arises from
the mismatch between the diagonalizations of $M^{}_e$ and
$M^{}_\nu$. On the other hand, it is possible to interpret the
cosmological baryon number asymmetry $\eta^{}_B \equiv
n^{}_B/n^{}_\gamma = (6.1 \pm 0.2) \times 10^{-10}$ \cite{WMAP}
with the help of the thermal leptogenesis mechanism \cite{FY}:
either through the out-of-equilibrium decay of the lightest
right-handed Majorana neutrino, or via the out-of-equilibrium
decay of one or more Higgs triplets \cite{TRI}, or due to both
effects. The left-right symmetric model is therefore an intriguing
playground to explore new physics beyond the standard model, at
least in the neutrino sector.

However, it is a highly non-trivial task to simultaneously account
for the cosmological baryon number asymmetry and current neutrino
oscillation data in the left-right symmetric model. The reason is
simply that the specific textures of $M^{}_e$, $M^{}_D$ and ${\cal
F}$ are not fixed by the model itself. To get around this
difficulty, one may impose certain flavor symmetries or empirical
assumptions on those mass matrices such that their textures can be
(partly) determined or constrained. Such a phenomenological
strategy has been adopted in some recent attempts
\cite{formula,leplr} to study neutrino mixing and leptogenesis
based on the left-right symmetry, although not all of them are
successful in fitting the updated observational \cite{WMAP} and
experimental \cite{Vissani} data.

The purpose of this paper is to propose two novel and viable
scenarios for $M^{}_D$ and ${\cal F}$ in the flavor basis where
$M^{}_e$ is diagonal, so as to simultaneously interpret the
observed neutrino mixing pattern and the observed baryon number
asymmetry of the Universe in the left-right symmetric model. A
salient feature of both scenarios is that the Higgs triplet Yukawa
coupling matrix takes the Friedberg-Lee (FL) texture
\cite{FL,FL2,FL3},
\begin{eqnarray}
{\cal F} \; =\; \left( \matrix{ b + c & -b & -c \cr -b & a + b &
-a \cr -c & -a & a + c \cr} \right) + d I \; ,
%---------------(5)
\end{eqnarray}
where $I$ denotes the $3\times 3$ identity matrix. Such a texture
is phenomenologically appealing for two simple reasons: (1) ${\cal
F}$ can be diagonalized by a unitarity transformation whose form
is very close to the interesting tri-bimaximal mixing pattern
\cite{TB}; and (2) the inverse matrix of ${\cal F}$, which appears
in the seesaw formula, has a structure exactly parallel to ${\cal
F}$. It is therefore possible to obtain the nearly tri-bimaximal
neutrino mixing matrix, which is particularly favored by current
neutrino oscillation data, from the Type-II seesaw relation under
a suitable condition. The complex phases of ${\cal F}$ turn out to
be the common source of CP violation for neutrino oscillations and
baryogenesis via leptogenesis, if the Dirac neutrino mass matrix
$M^{}_D$ is real. To be specific, we shall assume that $M^{}_D =
M^{}_e$ is diagonal and real in scenario (A), and $M^{}_D$ is real
and has a similar FL texture in scenario (B). We are going to
demonstrate that both scenarios are viable in the left-right
symmetric model to account for the cosmological baryon number
asymmetry and the neutrino mixing data.

The remaining part of this paper is organized as follows. In
section II, we diagonalize the FL texture and describe the picture
of leptogenesis as the preliminaries. Sections III and IV are
devoted to the details of scenarios (A) and (B), respectively. The
consequences of both scenarios on neutrino mixing and leptogenesis
are also illustrated in these two sections. A summary of our main
results is given in section V.

\section{Preliminaries}

In this section, we first describe a generic diagonalization of
the FL texture and then outline a couple of basic formulas to be
used for the calculation of leptogenesis.

\subsection{Diagonalization of ${\cal F}$}

The symmetric matrix ${\cal F}$ in Eq. (5) can be diagonalized by
the transformation $U^\dagger {\cal F} U^* = {\rm Diag} \{f^{}_1,
f^{}_2, f^{}_3 \}$, where $U$ is unitary and $f^{}_i$ (for
$i=1,2,3$) are real and positive. The special texture of ${\cal
F}$ guarantees $U$ to take the form
\begin{eqnarray}
U = \left ( \matrix{ \displaystyle \frac{2}{\sqrt{6}} &
\displaystyle \frac{1}{\sqrt{3}} & 0 \cr\cr \displaystyle
-\frac{1}{\sqrt{6}} & \displaystyle \frac{1}{\sqrt{3}} &
\displaystyle \frac{1}{\sqrt{2}} \cr\cr \displaystyle
-\frac{1}{\sqrt{6}} & \displaystyle \frac{1}{\sqrt{3}} &
\displaystyle -\frac{1}{\sqrt{2}} \cr } \right ) \left ( \matrix{
\displaystyle \cos{\theta} & 0 & \displaystyle \sin{\theta}
e^{-i\delta} \cr\cr 0 & 1 & 0 \cr\cr \displaystyle -\sin{\theta}
e^{i\delta} & 0 & \displaystyle \cos{\theta} \cr } \right ) \left
( \matrix{ e^{i\phi^{}_{1}} & 0 & 0 \cr\cr 0 & e^{i\phi^{}_2} & 0
\cr\cr 0 & 0 & e^{i\phi^{}_3} \cr } \right ) \; ,
%--------------------(6)
\end{eqnarray}
where the leading term is just the tri-bimaximal mixing pattern
\cite{TB}, and the parameters $\theta$ and $\delta$ are given by
\begin{eqnarray}
\tan 2\theta & = & \frac{2 \sqrt{\left (|A|^2 + |B|^2 \right )
|C|^2 + 2 {\rm Re}\left ( A^* B^* C^2 \right )}}{|B|^2-|A|^2} \; ,
\nonumber \\
\tan\delta & = & - \frac{{\rm Im}\left [ a^* \left( b - c \right)
+ b^*c \right ] }{{\rm Re}\left [ \left( a^* + d^* \right) \left(
b - c \right) \right ] +|b|^2-|c|^2} \; ,
%       (7)
\end{eqnarray}
together with $A = 3(b+c)/2 +d$, $B = 2a + (b+c)/2 + d$ and $C = -
\sqrt{3}(b-c)/2$ \cite{FL2}. Furthermore, we obtain
\begin{eqnarray}
f^{}_{1} & = & \left |A - C \tan\theta e^{-i\delta} \right | \; ,
\nonumber \\
f^{}_{2} & = & \left |d \right | \; ,
\nonumber \\
f^{}_{3} & = & \left |B + C \tan\theta e^{+i\delta} \right | \; ;
%       (8)
\end{eqnarray}
and
\begin{eqnarray}
\phi^{}_1 & = &  \frac{1}{2} \arg \left( A - C \tan\theta ~
e^{-i\delta} \right) \; ,
\nonumber \\
\phi^{}_2 & = &  \frac{1}{2}\arg \left( d \right) \; ,
\nonumber \\
\phi^{}_3 & = &  \frac{1}{2}\arg \left( B + C \tan\theta ~
e^{+i\delta} \right ) \; .
%       (9)
\end{eqnarray}
Without loss of generality, one may rotate away $\phi^{}_3$
through the redefinitions $\rho \equiv \phi_1^{} - \phi_3^{}$ and
$\sigma \equiv \phi_2^{} - \phi_3^{}$. Then $U$ contains three
non-trivial phase parameters: $\delta$, $\rho$ and $\sigma$.

Since $M^{}_L = v^{}_L {\cal F}$ and $M^{}_R = v^{}_R {\cal F}$
hold in the left-right symmetric model, their mass eigenvalues can
be given in terms of $f^{}_1$, $f^{}_2$ and $f^{}_3$ obtained in
Eq. (8). For example, three heavy right-handed Majorana neutrino
masses are simply $M^{}_i = v^{}_R f^{}_i$ (for $i=1,2,3$).

\subsection{Leptogenesis}

For simplicity, let us assume that three heavy right-handed
Majorana neutrinos have a normal mass hierarchy (i.e., $M^{}_1 <
M^{}_2 < M^{}_3$) and their masses are much smaller than the
masses of Higgs triplets. In this case only the CP-violating
asymmetry generated from the out-of-equilibrium decay of the
lightest right-handed Majorana neutrino can survive and contribute
to the thermal leptogenesis \cite{FY}. Such a CP-violating
asymmetry, denoted as $\varepsilon^{}_1$, arises from the
interference between the tree-level and one-loop decay amplitudes.
In the basis where $M_R^{}$ is diagonal and real,
$\varepsilon^{}_1$ is given by $\varepsilon^{}_1 =
\varepsilon^{(1)}_1 + \varepsilon^{(2)}_1$ with
\begin{eqnarray}
\varepsilon^{(1)}_1 & = & \frac{1}{8 \pi v^2} \sum_{j\not= 1}
\frac{{\rm Im} \left[\left(\hat M_D^\dagger \hat
M_D\right)^2_{1j}\right]}{\left(\hat M_D^\dagger \hat
M^{}_D\right)^{}_{11}} \sqrt{x^{}_j} \left[ \frac{2 - x^{}_j}{1 -
x^{}_j} - \left(1 + x^{}_j\right) \ln{\frac{x^{}_j + 1}{x^{}_j}}
\right] \; , \nonumber \\
\varepsilon^{(2)}_1 & = &  \frac{3}{16 \pi v^2} M^{}_{1}
\frac{{\rm Im}\left[\left(\hat M_D^\dagger \, M^{}_L \hat
M_D^*\right)_{11}\right]}{\left(\hat M_D^\dagger \hat
M^{}_D\right)_{11}} \; ,
%       (10)
\end{eqnarray}
where $v \simeq 174$ GeV, $x^{}_{j} = M^2_{j} / M^2_{1}$, and
$\hat{M}_D^{}$ is the Dirac neutrino mass matrix defined in the
basis where $M^{}_{R}$ is diagonal. Note that
$\varepsilon^{(1)}_1$ is the conventional CP-violating term, while
$\varepsilon^{(2)}_1$ is due to the one-loop contribution induced
by the Higgs triplets \cite{varep}.

The CP-violating asymmetry $\varepsilon^{}_1$ can give rise to a net
lepton number asymmetry in the Universe, and the nonperturbative
sphaleron interaction may partially convert this lepton number
asymmetry into a net baryon number asymmetry \cite{efficiency},
\begin{eqnarray}
\eta_B^{} \; \equiv \; \frac{n_B^{}}{n_\gamma^{}} \; \simeq \; - \;
0.96 \times 10^{-2} \varepsilon_1^{} \kappa_1^{} \; ,
%-----------(11)
\end{eqnarray}
where $\kappa^{}_1$ is an efficiency factor measuring the washout
effects associated with the out-of-equilibrium decay of the
lightest right-handed Majorana neutrino. The value of
$\kappa^{}_1$ can be evaluated from the following analytical
approximation \cite{efficiency}: $\kappa_1^{} = \kappa^-(K_1^{}) +
\kappa^+(K_1^{})$, where
\begin{eqnarray}
\kappa^-(K_1^{}) & = & -2 e^{-2 N(K_1^{})/3}
\left[ e^{2\overline{N}(K_1^{})/3} - 1 \right] \; , \nonumber \\
\kappa^+(K_1^{}) & = & {2\over z_B^{}(K_1^{}) K_1^{}} \left[ 1 -
e^{-2 z_B^{}(K_1^{}) K_1^{} \overline{N}(K_1^{})/3} \right] \; ,
%-----------(12)
\end{eqnarray}
with
\begin{eqnarray}
\overline{N}(K_1^{}) & = & {N(K_1^{})\over \left ( 1 +
\sqrt{\frac{N(K_1^{})}{N_{\rm eq}^{}}}\right )^2} \; ,
\nonumber \\
z_B^{}(K_1^{}) & \simeq & 1 + {1\over 2} \ln \left\{ 1 + {\pi
K_1^2 \over 1024} \left[\ln \left({3125\pi K_1^2 \over 1024}
\right) \right]^5 \right\} \; .
%-----------(13)
\end{eqnarray}
Here $K^{}_1$ is the ratio of the total decay width of the
lightest right-handed Majorana neutrino to the expansion rate of
the Universe at temperature $T = M^{}_{1}$ \cite{efficiency}. One
usually expresses $K^{}_1$ as $K^{}_1 = \tilde{m}^{}_1 /m^{}_*$,
where $\tilde{m}_1^{} = (\hat{M}_D^\dagger
\hat{M}^{}_D)_{11}^{}/M_{1}^{}$ denotes the effective (seesaw)
neutrino mass, and $m^{}_* \simeq 1.08 \times 10^{-3}$ eV stands
for the equilibrium neutrino mass. In addition,
$\overline{N}(K_1^{})$ represents the number density of the
lightest right-handed Majorana neutrino, which interpolates
between the maximal number densities $N_{\rm eq}^{} = 3/4$ and
$N(K_1^{}) = 9\pi K_1^{}/16$ for strong and weak washout regions,
respectively.

Note that the formulas listed above are only valid for the
flavor-independent leptogenesis mechanism. The flavor-dependent
effects will become relevant if thermal leptogenesis takes place
at temperatures below $M^{}_1 \sim 10^{12}$ GeV
\cite{Flavor,Antusch}. We shall assume $M^{}_1 > 10^{12}$ GeV for
two phenomenological scenarios to be discussed in the subsequent
sections, such that flavor effects on leptogenesis can be safely
neglected.

\section{Scenario (A)}

Scenario (A) is based on three assumptions: (1) $\kappa' \sim 0$
in Eq. (3), such that $M^{}_e \simeq \kappa \tilde{Y}^{}_l$ and
$M^{}_D \simeq \kappa Y^{}_l$ (with $\kappa \simeq v \simeq 174$
GeV) are both symmetric; (2) $Y^{}_l = \tilde{Y}^{}_l$ is
diagonal, or equivalently $M_D^{} = v {\rm Diag} \{y_e^{},
y_\mu^{}, y_\tau^{} \} = {\rm Diag} \{m_e^{}, m_\mu^{},
m_{\tau}^{} \}$; and (3) ${\cal F}$ takes the FL texture as shown
in Eq. (5). The first assumption is consistent with the general
expectation $\kappa' \ll \kappa$ in the left-right symmetric model
\cite{Ji}, while the second and third ones are purely
phenomenological. In these assumptions, the effective neutrino
mass matrix $M^{}_\nu$ in Eq. (4) can be explicitly expressed as
\begin{eqnarray}
M^{}_\nu & = & v^{}_{L} \left [ \left ( \matrix {b+c & -b & -c \cr
-b & a+b & -a \cr -c & -a & a+c \cr } \right ) + d I \right ] -
\frac{v^2}{v^{}_{R} S} \left [ \left ( \matrix { y_{e}^{2} \left(
\hat{b} + \hat{c} \right) & - y_{e}^{} y_{\mu}^{} \hat{b} & -
y_{e}^{} y_{\tau}^{} \hat{c} \cr - y_{e}^{} y_{\mu}^{} \hat{b} &
y_{\mu}^{2} \left( \hat{a} + \hat{b} \right) & - y_{\mu}^{}
y_{\tau}^{} \hat{a} \cr - y_{e}^{} y_{\tau}^{} \hat{c} & -
y_{\mu}^{} y_{\tau}^{} \hat{a} & y_{\tau}^{2} \left( \hat{a} +
\hat{c} \right) \cr } \right ) \right . \nonumber \\
& & \left . + ~ \frac{1}{d} \left ( \matrix{ y_e^2 & 0 & 0 \cr 0 &
y_\mu^2 & 0 \cr 0 & 0 & y_\tau^2 } \right ) \right ] \; ,
%---------------(14)
\end{eqnarray}
where $\hat{a}$, $\hat{b}$, $\hat{c}$ and $S$ are simple functions
of the parameters of $\cal F$:
\begin{eqnarray}
S & = & d \left[d^2 + 2 \left( a+b+c \right) d + 3 \left( ab+bc+ac \right)
\right] \; , \nonumber \\
\hat{a} & = & \frac{1}{S} \left[ -a d - \left( ab+bc+ac \right) \right] \; ,
\nonumber \\
\hat{b} & = & \frac{1}{S} \left[ -b d - \left( ab+bc+ac \right)
\right] \; ,
\nonumber \\
\hat{c} & = & \frac{1}{S} \left[ -c d - \left( ab+bc+ac \right)
\right] \; .
%        (15)
\end{eqnarray}
Note that the vevs of the Higgs bi-doublet and triplets are
related with one another through $v_L^{} v_R^{} = \gamma v^2$ in
the left-right symmetric model \cite{LR}, where $\gamma$ depends
on the Higgs potential of the model and is usually expected to be
${\cal O}(1)$. Typically choosing $v^{}_L \sim 0.1 {\rm eV}$
\cite{formula}, we may simplify Eq. (14) and get an approximate
expression of $M^{}_\nu$ as follows:
\begin{eqnarray}
M^{}_\nu \; \simeq \; v^{}_{L} \left [ \left ( \matrix{ b+c & -b &
-c \cr -b & a+b & -a \cr -c & -a & a+c \cr} \right ) + d I -
\Delta \left ( \matrix{ 0 & 0 & 0 \cr 0 & r^2 & -r \cr 0 & -r & 1
\cr } \right ) \right ] \; ,
%       (16)
\end{eqnarray}
where $r \simeq {y_\mu^{}/ y_\tau^{}}$ and $\Delta \sim
y_\tau^2(\hat{a} + d^{-1})/\gamma$. Because of $y^2_\tau \sim
10^{-4}$, the $\Delta$-term in Eq. (16) is strongly suppressed. It
is therefore a good approximation to take $M^{}_\nu \simeq M^{}_L$
(i.e., only the first two terms on the right-hand side of Eq. (16)
are kept). In this case, $M^{}_\nu$ has the FL texture and can be
diagonalized by using the unitary matrix $U$ given in Eq. (6). Three
light (left-handed) Majorana neutrino masses turn out to be $m^{}_i
= v^{}_L f^{}_i$ (for $i=1,2,3$), where $f^{}_i$ can be found from
Eq. (8). Comparing $m^{}_i$ with the masses of three right-handed
Majorana neutrinos $M^{}_i = v^{}_R f^{}_i$, we immediately arrive
at $M^{}_i/m_i^{} = v^{}_R/v^{}_L$ in scenario (A). Since $M^{}_1 <
M^{}_2 < M^{}_3$ has been required in discussing leptogenesis,
$m^{}_1 < m^{}_2 < m^{}_3$ must hold (i.e., the light Majorana
neutrinos have a normal mass hierarchy). Furthermore, three neutrino
mixing angles are given as
\begin{eqnarray}
\sin\theta^{}_{12} & = & \frac{1}{\sqrt{2 + \cos2\theta}} \; ,
\nonumber \\
\sin\theta^{}_{23} & = & \frac{\sqrt{2 + \cos2\theta -
\sqrt{3}\sin2\theta\cos\delta}}{\sqrt{2 \left (2 + \cos2\theta
\right )}} \; ,
\nonumber \\
\sin\theta^{}_{13} & = & \frac{2}{\sqrt{6}} \sin\theta \;
%       (17)
\end{eqnarray}
in the standard parametrization \cite{PDG06}. The CP-violating
phases of $U$ have been presented in Eqs. (7) and (9), from which
one may define $\rho = \phi_1^{} - \phi_3^{}$ and $\sigma =
\phi_2^{} - \phi_3^{}$ as two independent Majorana phases
\cite{FX01}. In view of $\theta^{}_{13} < 10^\circ$ as constrained
by a global analysis of current experimental data \cite{Vissani},
we obtain $\theta <12.2^\circ$ from Eq. (17). The smallness of
$\theta$ implies that $\sin\theta^{}_{12} \simeq 1/\sqrt{3}$ (or
$\theta^{}_{12} \simeq 35.3^\circ$) and $\sin\theta^{}_{23} \simeq
1/\sqrt{2}$ (or $\theta^{}_{23} \simeq 45^\circ$) are excellent
approximations; i.e., $U$ is a nearly tri-bimaximal neutrino
mixing pattern and is strongly favored by the solar and
atmospheric neutrino oscillation measurements.

Now let us consider leptogenesis in scenario (A). As one can see
from Eq. (10), the CP-violating asymmetry $\varepsilon_1^{}$
depends on the $(1, 1)$, $(1, 2)$ and $(1, 3)$ entries of
$\hat{M}_D^\dagger \hat{M}_D^{}$ as well as the $(1, 1)$ entry of
$\hat{M}_D^\dagger M_L^{}\hat{M}_D^*$, where $\hat{M}^{}_D =
M^{}_D U^*$ with $U$ being determined in Eq. (6). A
straightforward calculation yields
\begin{eqnarray}
\left(\hat{M}_D^\dagger \hat{M}_D^{}\right)_{11}^{} & \simeq &
m_\tau^2 \left (\frac{1}{3} - \frac{1}{6} \cos 2\theta -
\frac{1}{2\sqrt{3}} \sin2\theta \cos\delta \right ) \; , \nonumber \\
\left(\hat{M}_D^\dagger \hat{M}_D^{}\right)_{12}^{} & \simeq &
m_\tau^2 \left ( -\frac{1}{3\sqrt{2}} \cos\theta +
\frac{1}{\sqrt{6}} \sin\theta e^{i \delta} \right )
e^{i(\rho-\sigma)} \; ,\nonumber \\
\left(\hat{M}_D^\dagger \hat{M}_D^{}\right)_{13}^{} & \simeq &
m_\tau^2 \left ( \frac{1}{2\sqrt{3}} \cos^2\theta - \frac{1}{6}
\sin 2\theta e^{i \delta} - \frac{1}{2\sqrt{3}} \sin^2\theta e^{2i
\delta} \right ) e^{i\rho} \; ;
%----------(18)
\end{eqnarray}
and
\begin{eqnarray}
\left( \hat{M}_D^\dagger M^{}_L \hat{M}_D^{*} \right)_{11}^{} &
\simeq & m_\tau^2 e^{2i \left (\phi^{}_1 + \phi^{}_3 \right)}
\left[ m^{}_1 \left ( \frac{1}{6} \cos^2 \theta - \frac{1}{2
\sqrt{3}} \sin 2\theta e^{i\delta} + \frac{1}{2} \sin^2 \theta\
e^{2i\delta}
\right )^2 e^{2i\rho} \right .  \nonumber \\
& & \left. +  m^{}_2 \left ( \frac{1}{3 \sqrt{2}} \cos\theta -
\frac{1}{\sqrt{6}} \sin\theta e^{i\delta} \right )^2 e^{2i \sigma}
\right. \nonumber \\
& & \left. + m^{}_3 \left (\frac{1}{2 \sqrt{3}} \cos 2\theta +
\frac{1}{12} \sin 2\theta e^{-i\delta} - \frac{1}{4} \sin 2\theta
e^{i\delta} \right )^2 \right] \; ,
%          (19)
\end{eqnarray}
where the terms proportional to $m^2_e$ and those proportional to
$m^2_\mu$ have been omitted by taking account of $m^2_e \ll
m^2_\mu \ll m^2_\tau$. Eqs. (18) and (19) allow us to calculate
the CP-violating asymmetry $\varepsilon^{}_1$ via Eq. (10) and the
baryon number asymmetry $\eta^{}_B$ via Eqs. (11), (12) and (13)
in scenario (A).

For simplicity, we only consider a special but interesting
parameter space in our numerical analysis. We assume that $a$, $b$
and $c$ are real, $d$ is complex and $b=c$ holds. The physical
roles of different parameters in this simple example are rather
clear: (1) real $a$, $b$ and $c$ together with $b = c$ result in
the exact tri-bimaximal mixing \cite{TB}; (2) $m^{}_2 \simeq
v^{}_L |d|$ fixes the mass scale and spectrum of three left-handed
Majorana neutrinos; (3) the phase of $d$ is the only source of CP
violation which leads to non-vanishing $\varepsilon^{}_1$ and
$\eta^{}_{B}$; and (4) the $\Delta$-induced term in Eq. (16) is
essentially negligible. We generate the input points of those free
parameters by scanning their possible ranges according to a flat
random number distribution. Hence the output points will be a
clear reflection of the strong constraints, imposed by scenario
(A) itself and by current neutrino oscillation data, on relevant
parameters. The following experimental data have been taken into
account in our calculations: $30^\circ \leq \theta_{12} \leq
38^\circ$, $36^\circ \leq \theta_{23} \leq 54^\circ$ and
$\theta_{13} < 10^\circ$ as well as $\Delta m^2_{21} \equiv m^2_2
- m^2_1 = (7.2 \cdot\cdot\cdot 8.9) \times 10^{-5} ~{\rm eV}^2$
and $\Delta m^2_{32} \equiv m^2_3 - m^2_2 = \pm (2.1
\cdot\cdot\cdot 3.1) \times 10^{-3} ~{\rm eV}^2$ \cite{Vissani}.
We numerically demonstrate that this scenario can successfully
account for the cosmological baryon number asymmetry and have no
conflict with the neutrino oscillation measurements. Some results
are summarized below.
\begin{itemize}
\item       In FIG. 1 we show the predicted values of $\eta^{}_{B}$
changing with $M^{}_1$, the mass of the lightest right-handed
Majorana neutrino. We find that the observationally-allowed range
of $\eta^{}_B$ (i.e., $\eta^{}_B = (6.1 \pm 0.2) \times 10^{-10}$
\cite{WMAP}) can be reproduced from the flavor-independent
leptogenesis in the chosen parameter space with $4.9 \times
10^{12} ~ {\rm GeV} \leq M^{}_1 \leq 7.7 \times 10^{14} ~{\rm
GeV}$. In particular, $M^{}_1 \sim 10^{14}$ GeV is most favored.

\item       Taking $M^{}_1 = 10^{14}$ GeV for example,
we are able to fix $a = \pm (0.17 \cdots 0.33)$, $b = c = \pm
(0.023 \cdots 0.037)$, $|d| = (0.085 \cdots 0.099)$ and $\arg(d) =
\pm (1.0^{\circ} \cdots 18.3^{\circ})$. The values of these
parameters are not sensitive to the change of $M^{}_1$ in its
allowed range. It is worth remarking that the constraint on $|d|$
comes mainly from the choice $v^{}_L \sim 0.1$ eV and the
requirement $M^{}_1 < M^{}_2 < M^{}_3$, which is equivalent to
$m^{}_1 < m^{}_2 < m^{}_3$. On the other hand, $\arg(d)$ is
restricted by both $\eta^{}_{B}$ and the neutrino oscillation
data.

\item       Given $v^{}_R = 10^{16}$ GeV for instance, the
mass spectrum of three light Majorana neutrinos is $m^{}_1 = (0.55
\cdots 1.0) \times 10^{-3}$ eV, $m^{}_2 = (8.5 \cdots 9.8) \times
10^{-3}$ eV and $m^{}_3 = (4.2 \cdots 5.8) \times 10^{-2}$ eV; and
that of three heavy Majorana neutrinos is $M^{}_1 = (0.55 \cdots
1.0) \times 10^{14}$ GeV, $M^{}_2 = (8.5 \cdots 9.8) \times
10^{14}$ GeV and $M^{}_3 = (4.2 \cdots 5.8) \times 10^{15}$ GeV.
The normal hierarchy of $m^{}_i$ implies that the effective mass
of the neutrinoless double-beta decay $\langle m\rangle^{}_{ee}$
must be at the ${\cal O}(10^{-3})$ eV level \cite{Vissani}, which
is far below the present experimental upper bound $\langle
m\rangle^{}_{ee} < 0.35$ eV \cite{PDG06}.
\end{itemize}
Because of $b=c$ and the smallness of $\Delta$ taken in our
numerical calculations, the neutrino mixing matrix is essentially
the tri-bimaximal mixing pattern with a vanishingly small value of
$\theta^{}_{13}$. Hence there is no observable effect of CP
violation in neutrino oscillations. A detailed analysis of the FL
texture with $b \neq c$ can be found in Ref. \cite{FL2}. Here we
make a numerical check about the possible influence of $b\neq c$
on $\eta^{}_B$ in scenario (A). We find that the result shown in
FIG. 1 is actually not sensitive to the small difference between
$b$ and $c$.

\section{Scenario (B)}

Scenario (B) is based on three assumptions: (1) $\kappa' \sim 0$
in Eq. (3), such that $M^{}_e \simeq \kappa \tilde{Y}^{}_l$ and
$M^{}_D \simeq \kappa Y^{}_l$ (with $\kappa \simeq v \simeq 174$
GeV) are both symmetric; (2) $\tilde{Y}^{}_l$ is diagonal, but
$Y^{}_l$ takes the FL texture and its parameters are all real; and
(3) ${\cal F}$ takes the FL texture as given in Eq. (5). The first
and second assumptions allow us to write out $M^{}_D$ as follows:
\begin{eqnarray}
M^{}_{D} \; = \; v \left [ \left ( \matrix{ b'+c' & -b' & -c' \cr
-b' & a'+b' & -a' \cr -c' & -a' & a'+c' \cr } \right ) + d'I
\right ] \; .
%       (20)
\end{eqnarray}
As $a'$, $b'$, $c'$ and $d'$ are all assumed to be real, it is
easy to diagonalize $M_D^{}$ by using the transformation
$V'^\dagger M_D^{}V'^* = {\rm Diag} \{{D}^{}_{1}, {D}^{}_{2},
D^{}_3 \}$, where
\begin{eqnarray}
V^{'} \; = \; \left ( \matrix{ \displaystyle \frac{2}{\sqrt{6}} &
\displaystyle \frac{1}{\sqrt{3}} & 0 \cr\cr \displaystyle
-\frac{1}{\sqrt{6}} & \displaystyle \frac{1}{\sqrt{3}} &
\displaystyle \frac{1}{\sqrt{2}} \cr\cr \displaystyle
-\frac{1}{\sqrt{6}} & \displaystyle \frac{1}{\sqrt{3}} &
\displaystyle -\frac{1}{\sqrt{2}} \cr } \right ) \left ( \matrix{
\displaystyle \cos{\theta'} & 0 & \displaystyle \sin{\theta'}
\cr\cr 0 & 1 & 0 \cr\cr \displaystyle -\sin{\theta'} & 0 &
\displaystyle \cos{\theta'} \cr } \right ) \; .
%----------(21)
\end{eqnarray}
Similar to $\theta$ in Eq. (7), the rotation angle $\theta'$ can
also be determined in terms of $a'$, $b'$, $c'$ and $d'$. Taking
account of Eqs. (5) and (20), we calculate the effective Majorana
neutrino mass matrix $M^{}_\nu$ by means of the Type-II seesaw
formula in Eq. (4). We find that $M^{}_\nu$ has the FL texture as
${\cal F}$ and $M^{}_D$ do:
\begin{eqnarray}
M_\nu^{} \; = \; v_L^{} \left[\left(\matrix{ \tilde{b} + \tilde{c} &
-\tilde{b} & -\tilde{c} \cr - \tilde{b} & \tilde{a} + \tilde{b} &
-\tilde{a} \cr -\tilde{c} & -\tilde{a} & \tilde{a} + \tilde{c}}
\right) + \tilde{d} I \right] \; ,
%---------(22)
\end{eqnarray}
where
\begin{eqnarray}
\tilde{a} & = & -\left\{ \gamma d \left [ d^2 + 2 \left( a + b + c
\right) d + 3 \left( bc + ab + ac \right) \right ] \right \}^{-1}
\left \{ a' d \left( a' + d' \right ) \left [ 3 \left ( b + c \right )
+ 2d \right ]  \right. \nonumber\\
& & \left. + d \left [ a' b' \left ( 3c + d \right ) + a' c' \left
( 3b + d \right ) - b' c' \left ( 3a + d \right ) \right ] - d' d
\left [ b' \left ( a-c \right ) + c' \left ( a-b \right )
\right]  \right \} + a  \; , \nonumber\\
\tilde{b} & = & -\left\{ \gamma d \left [ d^2 + 2 \left( a + b + c
\right) d + 3 \left( bc + ab + ac \right) \right ] \right \}^{-1}
\left \{ b' d \left (b' + d' \right ) \left [ 3 \left ( a + c \right )
+ 2d \right ] \right. \nonumber\\
& & \left. + d \left [ a' b' \left ( 3c + d \right ) + b' c' \left (
3a + d \right ) - a' c' \left ( 3b + d \right ) \right ] - d' d
\left [ a' \left ( b-c \right ) + c' \left ( b-a \right )
\right]  \right \} + b  \; , \nonumber\\
\tilde{c} & = & -\left\{ \gamma d \left [ d^2 + 2 \left( a + b + c
\right) d + 3 \left( bc + ab + ac \right) \right ] \right \}^{-1}
\left \{ c' d \left (c' + d' \right ) \left [ 3 \left ( a + b \right )
+ 2d \right ] \right. \nonumber\\
& & \left. + d \left [ a' c' \left ( 3b + d \right ) + b' c' \left (
3a + d \right ) - a' b' \left ( 3c + d \right ) \right ] - d' d
\left [ a' \left ( c-b \right ) + b'
\left ( c-a \right ) \right]  \right \} + c \; , \nonumber\\
\tilde{d} & = & -\frac{{d '}^2}{\gamma d} + d  \;  .
%       (23)
\end{eqnarray}
It is obvious that $M^{}_\nu$ can be diagonalized by a unitary
transformation $V(\hat{\theta}, \hat{\delta}, \hat{\phi_1},
\hat{\phi_2}, \hat{\phi_3})$ which has the same form as $U$ given
in Eq. (6). One may simply use Eqs. (7), (8) and (9) to determine
the angle and phase parameters of $V$ as well as three neutrino
mass eigenvalues $m^{}_i$, after the replacements
$a\rightarrow\tilde{a}$, $b\rightarrow\tilde{b}$,
$c\rightarrow\tilde{c}$, $d\rightarrow\tilde{d}$ are made.

We proceed to consider leptogenesis in scenario (B). A noteworthy
feature of this scenario is that the $(1,2)$ entry of
$\hat{M}_D^\dagger \hat{M}_D^{}$ vanishes. The non-vanishing $(1,
1)$ and $(1, 3)$ elements of $\hat{M}_D^\dagger \hat{M}_D^{}$
together with the $(1, 1)$ element of $\hat{M}_D^\dagger
M_L^{}\hat{M}_D^{*}$ are given by
\begin{eqnarray}
\left( \hat{M}_D^\dagger \hat{M}_D^{} \right)_{11}^{} & = & D_1^2
\cos^2 \left ( \theta - \theta' \right ) + D_3^2 \sin^2 \left (
\theta - \theta' \right ) +\left ( D_3^2 - D_1^2 \right ) \sin
2\theta \sin 2\theta' \sin^2 \frac{\delta}{2}  \; , \nonumber\\
\left( \hat{M}_D^\dagger \hat{M}_D^{} \right)_{13}^{} & = &
\frac{1}{2} \left ( D_3^2 - D_1^2 \right ) \left [ \sin 2\theta
\cos 2\theta' - \sin 2\theta' \left ( \cos^2 \theta - \sin^2
\theta e^{2i\delta} \right ) \right ] e^{i \rho} \; ;
%       (24)
\end{eqnarray}
and
\begin{eqnarray}
\left( \hat{M}_D^\dagger M_L \hat{M}_D^{*} \right)_{11}^{} \; = \;
m^{}_1 T_1^2 + m^{}_3 T_3^2 \; ,
%       (25)
\end{eqnarray}
where
\begin{eqnarray}
T^{}_1 & = & \left [ D_1^{} \left ( \cos\theta \cos \theta' +
\sin\theta \sin\theta' e^{-i \delta} \right )^2 + D_3^{} \left (
\cos\theta \sin\theta' - \sin\theta \cos\theta' e^{i \delta}
\right )^2 \right ] e^{2i \phi^{}_1}  \; , \nonumber\\
T^{}_3 & = & \frac{1}{2} \left [ D_1^{} \left ( -\cos 2\theta \sin
2\theta' + \sin 2\theta \cos^2 \theta' e^{-i \delta} + \sin
2\theta \sin^2 \theta' e^{i \delta} \right ) \right. \nonumber\\
& & \left. + D_3^{} \left ( \cos 2\theta \sin 2\theta' - \sin
2\theta \cos^2 \theta' e^{i \delta} - \sin 2\theta \sin^2 \theta'
e^{-i \delta} \right ) \right ] e^{i \left ( \phi^{}_1 + \phi^{}_3
\right )} \; .
%       (26)
\end{eqnarray}
We remark that the CP-violating phases come from ${\cal F}$. In
addition, Eq. (24) shows that the $CP$-violating asymmetry
$\varepsilon^{(1)}_1$ directly depends on $\theta$ (from ${\cal
F}$), $\theta'$ and $D^2_3 - D^2_1$ (from $M^{}_D$). If these
three quantities are very small, $\varepsilon^{(1)}_1$ will be
strongly suppressed.

Now we carry out a numerical analysis of neutrino mixing and
leptogenesis in scenario (B), just for the purpose of
illustration. We assume that $a$, $b$ and $c$ are real but $d$ is
complex. Furthermore, we assume $a' = a$ and $b'=b=c'=c$ in order
to simplify the calculations. The only source of CP violation is
the phase of $d$, similar to the situation in scenario (A). It is
easy to see that $b'=b=c'=c$ leads to $\theta = \theta' = 0$, and
thus $\varepsilon^{(1)}_1$ vanishes. Non-zero $\eta^{}_B$ is
attributed to non-zero $\varepsilon^{(2)}_1$ in this scenario. Our
numerical results are consistent with both the observational data
on $\eta^{}_B$ \cite{WMAP} and the experimental data on two
neutrino mass-squared differences and three mixing angles
\cite{Vissani}. Some comments are in order.
\begin{itemize}
\item       In FIG. 2 we show the predicted values of $\eta^{}_{B}$
changing with $M^{}_1$. We see that the observationally-allowed
range of $\eta^{}_B$ can be reproduced from the flavor-independent
leptogenesis in the chosen parameter space with $6 \times 10^{12}
~ {\rm GeV} \leq M^{}_1 \leq 1 \times 10^{16} ~{\rm GeV}$, where
higher values of $M^{}_1$ have been cut off.

\item       Taking $M^{}_1 = 10^{14}$ GeV for example, we arrive at
the allowed regions for the parameters of ${\cal F}$ and $M^{}_D$:
$a = \pm (0.20 \cdots 0.28)$, $b = c = \pm (0.026 \cdots 0.049)$,
$|d| = (0.085 \cdots 0.097)$ and $\arg(d) = \pm (0.23^{\circ}
\cdots 5.7^{\circ})$ together with $d' = (0 \cdots 0.3)$.

\item       As a consequence of $b'=b=c'=c$, the neutrino mixing matrix
is exactly the tri-bimaximal mixing pattern with $\theta^{}_{12} =
35.3^\circ$, $\theta^{}_{23} = 45^\circ$ and $\theta^{}_{13} =
0^\circ$. Hence there is no CP violation in neutrino oscillations.
Given $v^{}_R = 10^{16}$ GeV for instance, the mass spectra of
light and heavy Majorana neutrinos are $m^{}_1 = (5.8 \cdots 9.7)
\times 10^{-4}$ eV, $m^{}_2 = (8.5 \cdots 9.5) \times 10^{-3}$ eV,
$m^{}_3 = (4.2 \cdots 5.8) \times 10^{-2}$ eV and $M^{}_1 = (5.8
\cdots 9.7) \times 10^{13}$ GeV, $M^{}_2 = (8.8 \cdots 9.7) \times
10^{14}$ GeV, $M^{}_3 = (4.3 \cdots 6.0) \times 10^{15}$ GeV,
respectively.
\end{itemize}
Again, the normal hierarchy of $m^{}_i$ implies that the effective
mass of the neutrinoless double-beta decay $\langle
m\rangle^{}_{ee}$ can only reach the ${\cal O}(10^{-3})$ eV level.

\section{Summary}

We have proposed two viable Type-II seesaw scenarios for the
neutrino mass matrix in the left-right symmetric model. The most
salient feature of our scenarios is that the Higgs triplet Yukawa
coupling matrix ${\cal F}$ takes the intriguing Friedberg-Lee
texture. In the basis where the charged-lepton mass matrix
$M^{}_e$ is diagonal, the Dirac neutrino mass matrix $M^{}_D$ has
been assumed to be identical to $M^{}_e$ in scenario (A) and to
take the FL texture in scenario (B). We have shown that the nearly
tri-bimaximal neutrino mixing pattern, which is particularly
favored by current neutrino oscillation data, can be naturally
derived from both scenarios. Requiring the lightest right-handed
Majorana neutrino mass $M^{}_1$ to be above $10^{12}$ GeV, we have
demonstrated the parameter space of each scenario in which the
cosmological baryon number asymmetry can be interpreted via the
flavor-independent leptogenesis mechanism.

It is worth emphasizing that some of the assumptions made for the
above Type-II seesaw scenarios are just for the sake of
simplicity. Hence they are not demanded in more general cases. For
example, one may discuss the flavor-dependent thermal leptogenesis
to account for the observed baryon number asymmetry of the
Universe by allowing $M^{}_1$ to be below $10^{12}$ GeV
\cite{Antusch}. One may also allow all the parameters of ${\cal
F}$ to be complex and independent, in order to generate an
experimentally appreciable value for the smallest neutrino mixing
angle $\theta^{}_{13}$ and to give rise to the observable effect
of CP violation in neutrino oscillations. In this sense we
conclude that our scenarios, which are associated with the
left-right gauge symmetry and its spontaneous breaking as well as
the FL flavor symmetry and its explicit breaking, have very rich
implications and consequences in neutrino phenomenology.

\begin{acknowledgments}
One of us (Z.Z.X.) would like to thank A.H. Chan and C.H. Oh for
warm hospitality at the National University of Singapore, where
this paper was written. We are also grateful to H. Zhang and S.
Zhou for useful discussions. Our research was supported in part by
the National Nature Science Foundation of China.
\end{acknowledgments}

\newpage

%%%%%%%%%%%%  Fig. 1  SA   %%%%%%%%%%%
\begin{figure}
\begin{center}
\vspace{-0.5cm}
\includegraphics[bbllx=2.2cm, bblly=6.0cm, bburx=12.2cm, bbury=16.0cm,%
width=7.9cm, height=7.9cm, angle=0, clip=0]{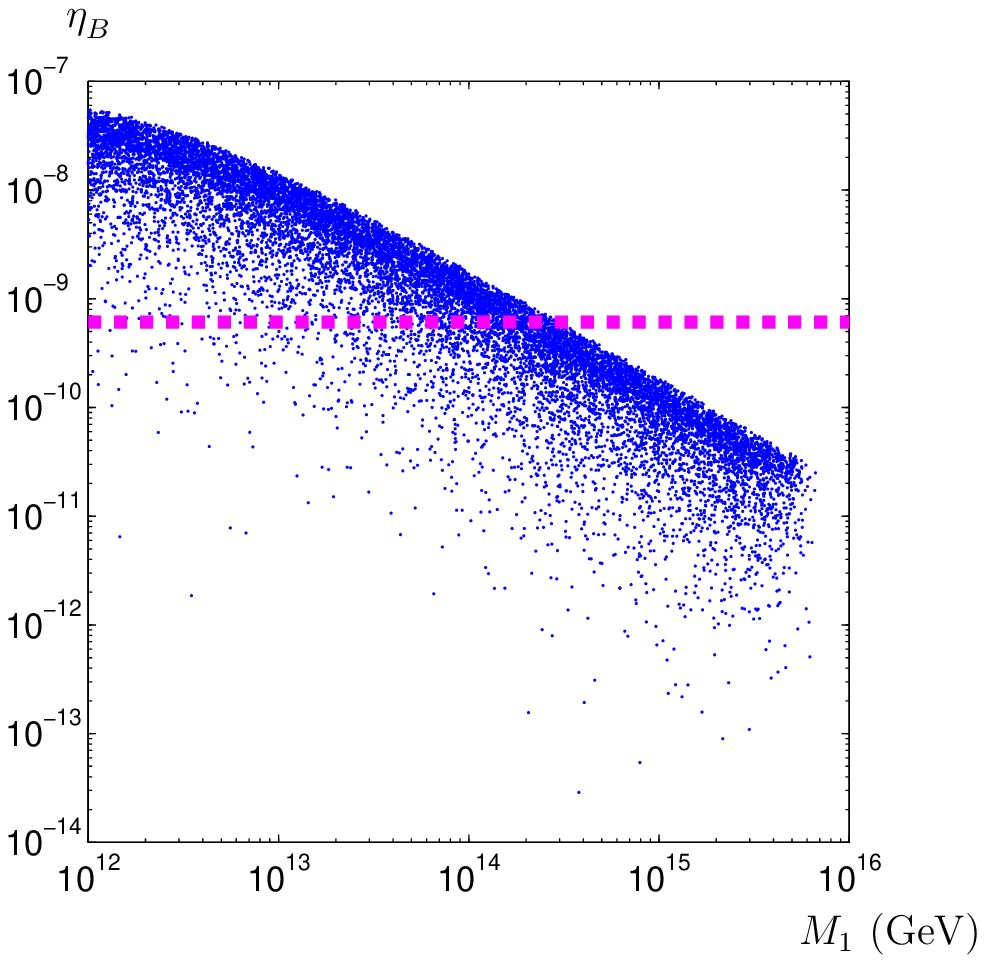}
\vspace{1.8cm}\caption{Illustrative plot for $\eta^{}_{B}$
changing with $M^{}_{1}$ in scenario (A). Here the dashed band
stands for the observationally-allowed range of $\eta^{}_{B}$.}
\end{center}
\end{figure}
%%%%%%%%%%%%%%%%%%%%%%%%%%%%%%%%%%%%%%%

%%%%%%%%%%%%  Fig. 3  SB   %%%%%%%
\begin{figure}
\begin{center}
\vspace{0cm}
\includegraphics[bbllx=2.2cm, bblly=6.0cm, bburx=12.2cm, bbury=16.0cm,%
width=7.9cm, height=7.9cm, angle=0, clip=0]{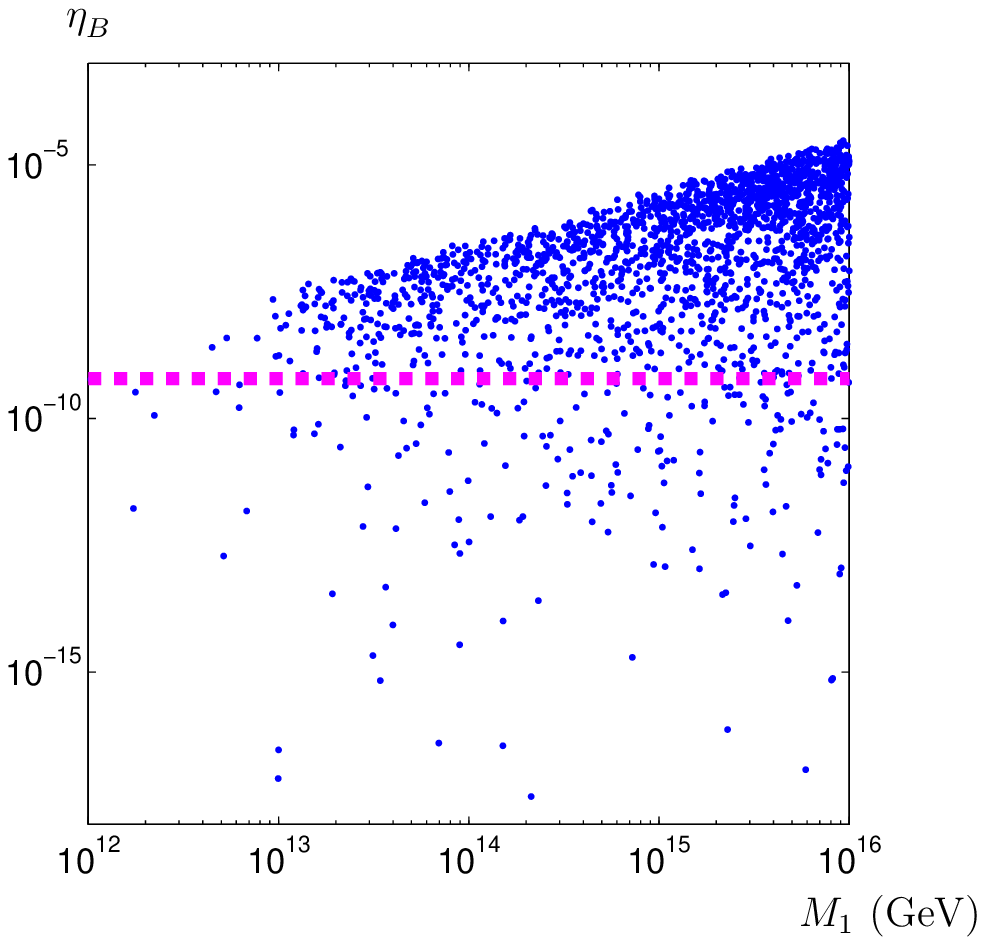}
\vspace{1.8cm}\caption{Illustrative plot for $\eta^{}_{B}$
changing with $M^{}_{1}$ in scenario (B). Here the dashed band
stands for the observationally-allowed range of $\eta^{}_{B}$.}
\end{center}
\end{figure}
%%%%%%%%%%%%%%%%%%%%%%%%%%%%%%%%%%%%%%%

\end{document}